\documentclass{appolb}

\usepackage{amsmath} 
\usepackage{graphicx}
\usepackage{diagbox}

\newcommand\MeV  {\ensuremath{\mathrm{MeV}}}
\newcommand\GeV  {\ensuremath{\mathrm{GeV}}}
\newcommand\TeV  {\ensuremath{\mathrm{TeV}}}
\newcommand\rc   {\ensuremath{\mathrm{c}}}

\newcommand\rs   {\ensuremath{\mathrm{s}}}

\newcommand\gL   {\ensuremath{g_\mathrm{L}}}

\newcommand\cw {\ensuremath{ \rc_W}}
\newcommand\sw {\ensuremath{ \rs_W}}

\newcommand\cz {\ensuremath{ \rc_Z}}
\newcommand\sz {\ensuremath{ \rs_Z}}

\newcommand\cs {\ensuremath{ \rc_s}}
\newcommand\sss {\ensuremath{ \rs_s}}

\begin{document}
\title{Precise prediction for the W boson mass in U(1) extensions of the standard model
\thanks{Presented at Matter to the Deepest 2023 international workshop.
This research was supported by the Excellence Programme of the Hungarian
Ministry of Culture and Innovation under contract TKP2021-NKTA-64.
}
}
\author{Zoltán Péli
\address{Institute for Theoretical Physics, ELTE Eötvös Loránd University,
\\Pázmány Péter sétány 1/A, 1117 Budapest, Hungary}
}
\maketitle
\begin{abstract}
We present the one-loop correction to the $W$ boson mass in U(1)$_z$ type extensions of the standard model. We compare it to an approximation, often used in high energy physics tools. We point out that if the $Z'$ boson -- predicted in U(1)$_z$ type extensions -- is much heavier than the $Z$ boson, then the use of the complete set of one-loop corrections is necessary.
\end{abstract}
  
\section{Introduction}
So far no new elementary particles have been observed besides the ones predicted by the standard model (SM). This is in contrast with the fact that we observed phenomena which cannot be explained in terms of the SM, although a particle physics origin for these phenomena is well motivated. Z. Trócsányi's contribution to this proceedings \cite{mttd23:tz} gives a detailed account on the beyond standard model (BSM) phenomena and their possible resolution in terms of a particular U(1)$_z$ type extension. These U(1)$_z$ type extensions are possibly the simplest realizations of a fifth fundamental force in nature, and hence a new gauge boson, denoted with $Z'$.

The non-observation of new BSM particles motivates
one to examine indirect effects of BSM physics in observable quantities. The excellent and ever increasing precision of experimental results and SM theory predictions allows one to expose quantum corrections which cannot be explained in terms of the SM if the deviation between the experimental result and theory prediction is sufficiently large.

Here we present the complete computation of one-loop corrections to the mass $M_W$ of the W boson in U(1)$_z$ extensions and argue that the precision of the BSM predictions is also an important feature, since it could give a false positive or a false negative prediction for the presence of new physics.

\section{W boson mass}
The $W$ boson is the charged mediator of the weak interaction. Both the experimental results and theoretical predictions for its mass have a high precision, with an uncertainty of about $10~\MeV$ or one part in ten thousand. This precision has the potential to expose BSM effects or to severely constrain the parameter space of a given model describing BSM physics.

\begin{figure}[htb]
\centerline{
\includegraphics[width=6.25cm]{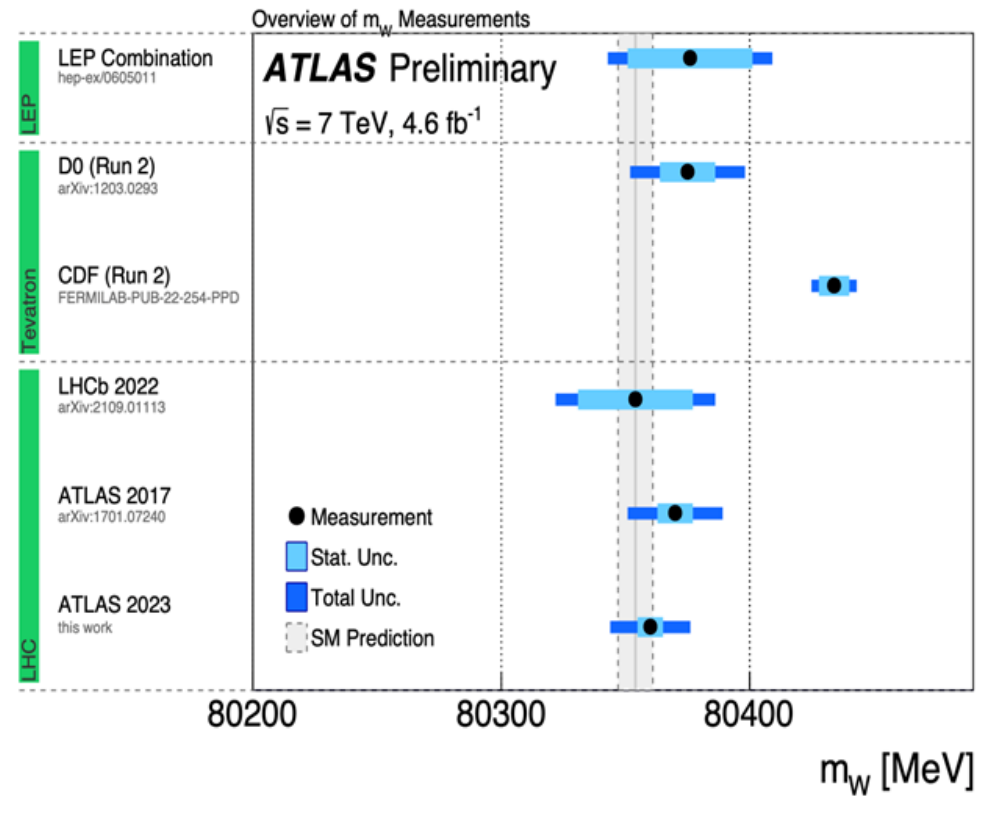}}
\caption{ATLAS Experiment: Measurements of the $W$ boson mass}
\label{fig:w_compilation}
\end{figure}

On the experimental side the world average cited in the 2022 version of the Particle Data Group \cite{Workman:2022ynf} is
\begin{equation}
    M_W^{\rm exp.} = \bigl( 80377 \pm 12 \bigr) ~\MeV.
    \label{eq_mw_exp}
\end{equation}
The $W$-mass gained a lot of attention after the CDFII result \cite{CDF:2022hxs}
was published, which is not yet included in the PDG world average. This motivated the ATLAS experiment at the Large Hadron Collider to reanalyze its 2011 data set and publish an improved result recently. The two results
\begin{equation}
     M_W^{\rm CDFII} = \bigl( 80434 \pm 9 \bigr)~\MeV
     \quad\text{and}\quad
     M_W^{\rm ATLAS} = \bigl( 80360 \pm 16 \bigr)~\MeV.
\end{equation}
differ by four standard deviations.
A compilation of experimental results by the ATLAS experiment is shown in Fig.~\ref{fig:w_compilation}.

On the theory side, $M_W$ is computed from the decay width of the muon particle as 
\begin{equation}
    \frac{G_F}{\sqrt{2}} = \frac{\pi\alpha}{2 M_W^2 \sw^2}\bigl(1+\Delta r\bigr), 
    \quad \textrm{with} \quad
    \cw = \frac{M_W}{M_Z},
\end{equation}
where $G_F$ is the Fermi constant, $\alpha$ is the fine-structure constant, $M_Z$ is the mass of the $ Z$ boson and finally $\Delta r$ collects the quantum corrections. The sine and cosine of the weak mixing angle $\theta_W$ is abbreviated as $\sw$ and $\cw$ and we adopt this shorthand for other mixing angles in the rest of this proceedings. The $\Delta r$ at one-loop in the SM was first computed in 1980 \cite{Sirlin:1980nh} and it is given as 
\begin{equation}
\begin{split}
\Delta r_{\rm SM} &= 
\frac{2 \delta e}{e}
+\frac{\text{Re}\Pi_{WW}(M_W^2)-\Pi_{WW}(0)}{M_W^2}+\delta_{\text{BV}}
\\&
\quad
+\frac{\cw^2}{\sw^2 M_W^2}\biggl[
\cw^2\text{Re}\Pi_{ZZ}(M_Z^2)
-\text{Re}\Pi_{WW}(M_W^2)\biggr]
\label{eq:deltar-os}
\end{split}
\end{equation}
in the on-shell renormalization scheme, where $\delta e$ is the counterterm for the electric charge, $\delta_{\text{BV}}$ are the box and vertex corrections to the muon decay process and $\Pi_{VV}(p^2)$ is the transverse piece of the one-loop self-energy of the vector boson $V=Z,W$. 
Currently, the complete set of two-loop corrections are computed to $\Delta r$ with leading three-loop corrections \cite{Awramik:2002vu,Degrassi:2014sxa}. We evaluate $M_W$ according to the fit formula of Ref.~\cite{Degrassi:2014sxa} using input parameters taken from Ref.~\cite{Workman:2022ynf} to obtain the SM prediction
\begin{equation}
    M_W^{\rm theo.} = \bigl( 80353 \pm 9 \bigr)~\MeV,
    \label{eq:mw_sm}
\end{equation}
where the largest sources of the uncertainty are the uncertainties of the top quark mass and the hadronic vacuum polarization.

\section{U(1)$_z$ extensions and $M_W$}
The U(1)$_z$ extensions are relatively simple extensions of the SM with the potential to explain several observed BSM phenomena. The model we consider here contains a new neutral, massive or massless gauge boson $Z'$, a new complex scalar field $\chi$ and right-handed sterile neutrinos. The idea is that the new scalar field obtains a vacuum expectation value (VEV), which spontaneously breaks the U(1)$_z$ symmetry and provides mass to the neutrinos. The charge assignment is constrained by the requirement to cancel gauge and gravity anomalies and to have gauge invariant Yukawa interactions, which leaves the $z$-charge of two fields unconstrained. This gives rise to several, phenomenologically different U(1)$_z$ extensions, such as in 
Refs.~\cite{Schabinger:2005ei,Pospelov:2007mp,Basso:2008iv,Trocsanyi:2018bkm}.
and many others.

In order to streamline our discussion we neglect the effect of the sterile neutrinos in the following. In this case, there are five free parameters, for which we select phenomenoligically useful ones, i.e. 
\begin{itemize}
    \item the VEV $w$ of the new scalar field, or rather its ratio to the VEV $v$ of Brout-Englert-Higgs field $\phi$  as $\tan\beta = w/v$
    \item the mass of the new scalar boson $M_s$ and the sine $\sss$ of the scalar mixing angle $\theta_s$ used to diagonalize the mass matrix of the scalar particles 
    \begin{equation}
    \left(\begin{array}{c}
    \phi^0 \\
    \chi
    \end{array}\right)  =
    \left( \begin{array}{cc}
    \cs & -\sss  \\
    \sss &  \cs  \end{array}
    \right)
     \left(\begin{array}{c}
    h \\
    s
    \end{array}\right),
    \end{equation}
    where the fields on the right hand side are the mass eigenstates.
    \item the mass of the new gauge boson $M_{Z'}$ and the sine $\sz$ of the mixing angle $\theta_Z$ used to diagonalize the mass matrix of the neutral gauge bosons together with the weak mixing angle 
    \begin{equation}
    \left(\begin{array}{c}
    B_\mu \\
    W^{3}_\mu \\
    B'_\mu
    \end{array}\right) =
    \left( \begin{array}{ccc}
    \cw & -\sw & 0 \\
    \sw &  \cw & 0 \\
      0 &  0   &  1 \end{array}
    \right)
    \left( \begin{array}{ccc}
     1 &   0 &  0   \\
    0 & \cz & -\sz \\
    0 & \sz &  \cz \end{array}
    \right)
    \left(\begin{array}{c}
    A_\mu \\
    Z_\mu \\
    Z^{\prime}_\mu
    \end{array}\right),
    \label{eq:gauge-rotation}
    \end{equation}
    where again the fields on the right hand side are the mass eigenstates and $B_\mu'$ is the new U(1) gauge field.
\end{itemize}

The masses $M_Z$ and $M_{Z'}$ and also the mixing angle $\theta_Z$ are complicated expressions of the Lagrangian couplings, but these can be written in compact forms 
\begin{equation}
\begin{split}
    \tan(2 \theta_Z) &= -\frac{2 \kappa}{1 - \kappa^2 - \tau^2},\\
    M_Z = \frac{M_W}{c_W}\sqrt{R(\cz,\sz)}, \quad&\quad
    M_Z = \frac{M_W}{c_W}\sqrt{R(\cz,\sz)},
\end{split}
\label{eq:MZMZpM_mass}
\end{equation}
with 
\begin{equation}
    M_W = \frac{1}{2}\gL v,\quad
    \cw = \frac{\gL}{\sqrt{g_y^2 + \gL^2}},\quad
    \text{and}\quad R(x,y) = (x- \kappa y)^2 + (\tau y)^2,
\end{equation}
where $g_y$ and $\gL$ are the U(1)$_Y$ and SU(2)$_L$ gauge couplings, $\kappa$ and $\tau$ are effective couplings composed of those Lagrangian couplings and $\tan\beta$. The mass matrix of the neutral Goldstone bosons can also be diagonalized in $R_\xi$ gauge, but it is not independent of the neutral gauge bosons as we expect that the mass squared of the Goldstone bosons are $\xi_Z M_Z^2$ and $\xi_{Z'} M_{Z'}^2$. After some cumbersome but straightforward algebra of the diagonalization we obtain the equations
\begin{equation}
    M_{Z'}\bigl(\cz  - \kappa \sz \bigr) = M_Z \cz \tau \quad\text{and}\quad
    M_Z \bigl(\sz  + \kappa \cz \bigr) = M_{Z'}\sz \tau,
\end{equation}
which can be solved for $\kappa$ and $\tau$ and substituted into \eqref{eq:MZMZpM_mass} to obtain
\begin{equation}
    \frac{M_W^2}{\cw^2} = M_Z^2 \cz^2 + M_{Z'}^2 \sz^2.
    \label{eq:MWCW}
\end{equation}
This formula shows us that the $\rho$-parameter is already modified at tree level by the U(1)$_z$ extension as
\begin{equation}
    \rho =  \frac{M_W^2}{M_Z^2\cw^2} = 1 - \sz^2 \biggl(1 - \frac{M_{Z'}^2}{M_Z^2} \biggr),
\label{eq:u1_rho}
\end{equation}
which in turn modifies the prediction for $M_W$ already at the tree level:
\begin{equation}
    M_W^2 = \frac{\rho M_Z^2}{2}\biggl(1  + \sqrt{1- \frac{4\pi \alpha}{\sqrt{2}G_F \bigl(\rho M_Z^2\bigr)}\bigl(1+\Delta r\bigr)} \biggr).
\end{equation}
The derivation of the radiative corrections $\Delta r$ is detailed in Ref.~\cite{Peli:2023fyb}, here we only highlight the fact that the renormalization of the weak mixing angle is different than in the SM.  If we split bare parameters $g^{(0)}$ into renormalized ones plus counterterms as $g^{(0)} = g + \delta g$, we see that counterterm $\delta\cw^2$ obtained from \eqref{eq:MWCW} is
\begin{equation}
    M_W^2 \frac{\delta\cw^2}{\cw^2} = \delta M_W^2 - \cw^2 \biggl(\cz^2 \delta M_Z^2 + \sz^2 \delta M_{Z'}^2 - 2\sz \bigl(M_Z^2 -M_{Z'}^2 \bigr)\delta\sz \biggr),
\label{eq:wrenorm}
\end{equation}
which is quite different than the one obtained in the SM, i.e in the $\theta_Z \to 0$ limit. The complete expression for $\Delta r$ at one loop in the on-shell scheme is derived in Ref.~\cite{Peli:2023fyb} and it is given as
\begin{equation}
\begin{split}
&\Delta r = 
\frac{2 \delta e}{e}
+\frac{\text{Re}\Pi_{WW}(M_W^2)-\Pi_{WW}(0)}{M_W^2}+\delta_{\text{BV}}
\\&
\quad
+\frac{\cw^2}{\sw^2 M_W^2}\biggl[
\cw^2\text{Re}\Pi_{ZZ}(M_Z^2)
-\text{Re}\Pi_{WW}(M_W^2)\biggr]
\\&
\quad
-\sz^2\frac{\cw^2}{\sw^2}\frac{\cw^2}{M_W^2}\biggl[
\text{Re} \Pi_{ZZ}(M_Z^2)
- \text{Re} \Pi_{Z'\!Z'}(M_{Z'}^2)
+2\bigl(M_Z^2\!-\!M_{Z'}^2\bigr)\frac{\delta \sz}{\sz}
\biggr].
\label{eq:deltarBSM}
\end{split}
\end{equation}
Upon comparison to \eqref{eq:deltar-os} one can recognize that the first two lines of \eqref{eq:deltarBSM} are formally the same as \eqref{eq:deltar-os}, but the self-energies and box and vertex diagrams in \eqref{eq:deltarBSM} also include loops with BSM particles and couplings. The counterterm for the new mixing angle can be computed using the relationships between field renormalization constants. The definition \eqref{eq:gauge-rotation} yields the relation between the unrotated fields and mass eigenstates both for the bare and the renormalized fields, 
\begin{equation}
\begin{split}
    B_\mu^{(0)} = \cw^{0} A_\mu^{(0)} - \sw^{(0)}\bigl(\cz^{(0)}Z_\mu^{(0)} - \sz^{(0)} Z_\mu^{'(0)} \bigr),&\quad
    B_\mu^{'(0)} =  \sz^{(0)}Z_\mu^{(0)} + \cz^{(0)} Z_\mu^{'(0)}
    \\
    B_\mu = \cw A_\mu - \sw\bigl(\cz Z_\mu - \sz Z_\mu' \bigr),&\quad
    B_\mu' =  \sz Z_\mu + \cz Z_\mu'.
\end{split}
\label{eq:field_1}
\end{equation}
The bare and renormalized unrotated fields are related as
\begin{equation}
    B_\mu^{(0)} = \sqrt{Z_B} B_\mu \quad\text{and}\quad
    B_\mu^{'(0)} = \sqrt{Z_{B'}} B_\mu',
\label{eq:field_2}
\end{equation}
wheres the mass eigenstates may mix as
\begin{equation}
    \left(\begin{array}{c}
    A_\mu^{(0)} \\
    Z_\mu^{(0)} \\
    Z_\mu^{'(0)}
\end{array}\right) =
\left( \begin{array}{ccc}
  \sqrt{Z_{AA}} &   \frac{1}{2} Z_{AZ} &  \frac{1}{2} Z_{AZ'}   \\
  \frac{1}{2} Z_{ZA} & \sqrt{Z_{ZZ}} & \frac{1}{2} Z_{ZZ'} \\
  \frac{1}{2} Z_{Z'A} & \frac{1}{2} Z_{Z'Z} &  \sqrt{Z_{Z'Z'}} \end{array}
\right)
\left(\begin{array}{c}
    A_\mu \\
    Z_\mu \\
    Z^{\prime}_\mu
\end{array}\right).
\label{eq:field_3}
\end{equation}
Combining eqs. \eqref{eq:field_1}, \eqref{eq:field_2} and \eqref{eq:field_3} yields the following relations
\begin{equation}
\begin{split}
    \sqrt{Z_B} \cw &= \cw^{(0)}\sqrt{Z_{AA}} - \frac{1}{2}\sw^{(0)}\bigl (\cz^{(0)}Z_{ZA} - \sz^{0} Z_{Z'A} \bigr),\\
    \sqrt{Z_{B'}}\sz &= \sz^{(0)}\sqrt{Z_{ZZ}} + \frac{1}{2}\cz^{(0)} Z_{Z'Z},\\
    \sqrt{Z_{B'}}\cz &= \frac{1}{2}\sz^{(0)} Z_{ZZ'} + \cz^{(0)}\sqrt{Z_{Z'Z'}}.
\end{split}
\label{eq:field_4}
\end{equation}
The first equation in \eqref{eq:field_4} yields the charge renormalization counterterm $\delta e$, once the U(1) Ward identity $Z_B Z_{g_y} = 1$ is used and the last two equations can be divided to cancel $Z_{B'}$ and to obtain the formula between the bare and renormalized mixing angle $\sz$ and hence providing $\delta \sz$. Let us remark here that \eqref{eq:field_4} is valid at all orders in perturbation theory and the explicit form of $\delta e$ is exactly the same as in the SM at one loop. 
The field renormalization constants can be obtained once the renormalization scheme is set, for instance in the on-shell scheme. 

We verified that the expression \eqref{eq:deltarBSM} is finite and independent of the gauge parameters in the $R_\xi$-gauge with a general charge assignment for the $z$-charges. Thus we obtain the formula for $\Delta r$ to compute how a U(1)$_z$ extension affects the mass of the W boson.

\section{Numerical analysis}

There are high energy physics tools, that are indispensable in an efficient and detailed analysis of a BSM model. Such tools are \texttt{SARAH/SPheno} \cite{Porod:2003um,Porod:2011nf,Staub:2009bi,Staub:2013tta} and \texttt{FlexibleSUSY} \cite{Athron:2014yba,Athron:2022isz}. In order to be as general as possible these tools do not consider new counterterms in the renomalization of the Weinberg angle (such as $\delta M_{Z'}^2$ and $\delta \sz$ in Eq.~\eqref{eq:wrenorm}) and hence the $\Delta r$ obtained in them does not contain the third line in Eq.~\eqref{eq:deltarBSM}. We investigate if there are regions of the parameter space, where the difference between the prediction for $M_W$ obtained using Eq.~\eqref{eq:deltarBSM} (case i) and using the truncation of the HEP tools for $\Delta r$ (case ii) becomes relevant.

In order to perform a numerical analysis we select the superweak extension of the SM \cite{Trocsanyi:2018bkm}, i.e the $z$-charges are $z_Q = 1/6$ for the left handed quark doublet and $z_U = 7/6$ for the right handed up quark. 
We also fix the gauge to the Feynman gauge, since the predictions for case ii depend on the gauge parameters.

The new gauge mixing angle $\theta_Z$ has to be rather small $\bigl(\sz \ll 1\bigr)$ in order to produce predictions consistent with experimental bounds. The dependence on $\sz$ and $M_{Z'}$ is more severe than on $\sss$ and $M_s$, which is not a surprise, since the $\rho$-parameter \eqref{eq:u1_rho} has a tree level dependence on $\sz$ and $M_{Z'}$. If the $Z'$ boson is lighter (heavier) than the $Z$ boson then it makes the $W$ boson lighter (heavier) than in the SM. We also remark here, that a very heavy $Z'$, i.e. $M_{Z'} \gg M_Z$ requires the new VEV to be sufficiently large $\bigl(M_{Z'}/w \sim O(1)\bigr)$, otherwise the new gauge couplings are not perturbative.

\begin{figure}[htb]
\centerline{
\includegraphics[width=6.25cm]{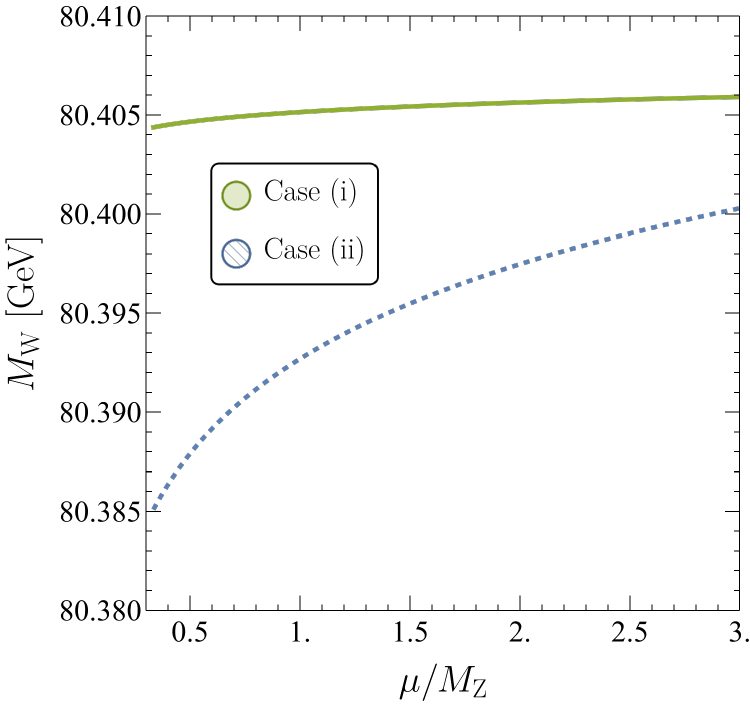}
\includegraphics[width=5.75cm]{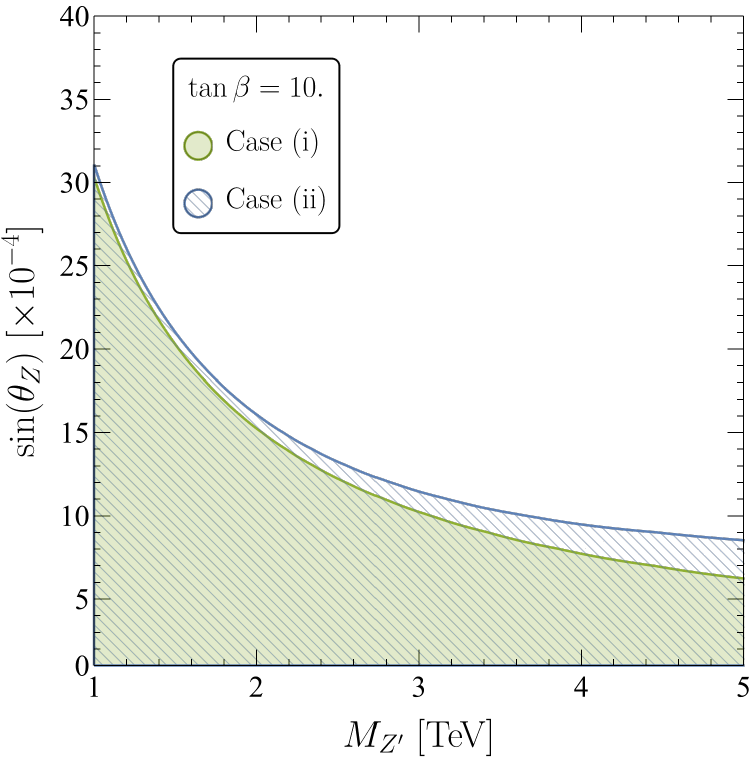}
}
\caption{On the left: The dependence of $M_W$ on the renormalization scale $\mu$ for the input parameters $M_{Z'} =3~\TeV$, $\sz = 10\cdot10^{-4}$, $\tan\beta=10$, $M_s = 700~\GeV$ and $\sss = 0.2$. On the right: The green and blue regions show the allowed values of $M_{Z'}$ and $\sz$ for $M_s = 700~\GeV$ and $\sss=0.2$.}
\label{fig:w1}
\end{figure}

We observe that the use case ii when $M_{Z'}\ll M_Z$ is completely justified.
In this region, the correction to $M_{W,\text{SM}}$ given in Eq.~\eqref{eq:mw_sm} is generally a few $\MeV$-s and the difference between the two cases is even smaller.
\begin{table}[t!]
\centering
\begin{tabular}{|l || c | c || c | c || c | c || c | c |} 
\hline
\hline
 $~\quad s_Z$ & \multicolumn{4}{|c||}{$5\cdot 10^{-4}$} & \multicolumn{4}{|c|}{$10\cdot 10^{-4}$}\\
\hline
\diagbox{$\tan\beta\!\!\!\!$}{$\!\!\!\!M_s$} 
& \multicolumn{2}{|c||}{$\begin{array}{c}0.7\,\TeV\\\text{(i)~~~(ii)}\end{array}$}
& \multicolumn{2}{|c||}{$\begin{array}{c}3\,\TeV\\\text{(i)~~~(ii)}\end{array}$} 
& \multicolumn{2}{|c||}{$\begin{array}{c}0.7\,\TeV\\\text{(i)~~~(ii)}\end{array}$}
& \multicolumn{2}{|c|}{$\begin{array}{c}3\,\TeV\\\text{(i)~~~(ii)}\end{array}$}
\\
\hline
\hline
10 & ~10~ & 6 & ~4~ & 3 & ~52~ & 40 & ~42~ & 40 \\
\hline 
20 & ~9~ & 8 & ~3~ & 3 & ~53~ & 50 & ~42~ & 45 \\ 
\hline
30 & ~9~ & 9 & ~3~ & 3 & ~53~ & 52 & ~43~ & 46\\ 
\hline
\hline
\end{tabular}
\caption{Predictions for $\Delta M_W = M_W - M_{W,{\rm SM}}$ in MeV units at parameter values $M_{Z'}=3~\TeV$ and $\sss=0.2$.}
\label{table:p2b}
\end{table}
We find that for a heavy $Z'$ $\bigl(M_{Z'}\gg M_Z\bigr)$ there are regions in the parameter space, where the difference between the predictions produced by cases i and ii is above $10~\MeV$, the present theory and experimental uncertainty of $M_W$. 
Table~\ref{table:p2b} shows benchmark points in the parameter space for a heavy $M_{Z'} = 5~\TeV$. 
The entries of the table are the pure BSM corrections in $\MeV$ units to the SM prediction $M_{W,\text{SM}}$ for several different parameters in both cases i and ii.
The top left entry of Table~\ref{table:p2b} shows a $27~\MeV$ deviation between the predictions for $M_W$ in cases i and ii. The renormalization scale dependence of these predictions is shown in the left plot of Fig.~\ref{fig:w1}. This shows us that the scale dependence of $M_W$ in case ii is more prominent than in case i.

Finally, we present a $2\sigma$ exclusion band for the parameter plane spanned by a heavy $M_{Z'}$ and $\sz$ based on the deviation between the experimental result \eqref{eq_mw_exp} and theoretical prediction for $M_W$
\begin{equation}
    \bigl|M_W^{\text{exp.}} - M_W\bigr| < 2 \sigma,
    \quad\text{with}\quad
    \sigma = \sqrt{\sigma^2_{\text{exp.}}+\sigma^2_{\text{theo.}}} = 15~\MeV,
    \label{eq:exclusion}
\end{equation}
where $M_W$ is computed using both cases i and ii.
The allowed region is shown in the right plot of Fig.~\ref{fig:w1} and reveals that for a heavy  $Z'$ the exclusion is more severe using the complete set of radiative corrections, i.e. case i. The difference between the bands becomes more pronounced with increasing values of $M_{Z'}$ for fixed $\tan\beta$, which also corresponds to increasing values of the new Lagrangian gauge couplings.

\section{Conclusion and outlook}

We presented the complete one-loop correction to the mass of the $W$ boson in U(1)$_z$ extensions, which was not available before. The prediction for $M_W$ using \eqref{eq:deltarBSM} (case i) is compared to a prediction for $M_W$ using a truncated set of one-loop corrections (case ii) used in HEP tools. We find that for a light $Z'$ boson, the truncation used in case ii is completely justified. In the region of the parameter space, where $M_{Z'}\gg M_Z$ the difference between cases i and ii may be large, hence using case i to compute $M_W$ is necessary. 
This proceedings is based on Ref.~\cite{Peli:2023fyb}, where a detailed account of the numerical analysis can be found. We plan to extend this work to compare the exclusion \eqref{eq:exclusion} using \eqref{eq:deltarBSM} to exclusion limits produced by direct searches in particle colliders.

\end{document}